\documentclass[prd,preprintnumbers,floatfix,aps,nofootinbib]{revtex4-1}

\usepackage{latexsym}
\usepackage{epsfig}
\usepackage{amssymb}
\usepackage{amsmath}
\usepackage[T1]{fontenc}
\usepackage[utf8]{inputenc}

\newcommand{\lp}{\left(}
\newcommand{\rp}{\right)}
\newcommand{\lb}{\left[}
\newcommand{\rb}{\right]}

\newcommand{\ba}{\begin{eqnarray}}
\newcommand{\ea}{\end{eqnarray}}
\newcommand{\be}{\begin{equation}}
\newcommand{\ee}{\end{equation}}

\newcommand{\om}{\omega}

\newcommand{\al}{\alpha}
\newcommand{\bt}{\beta}
\newcommand{\ga}{\gamma}
\newcommand{\ka}{\kappa}
\newcommand{\ta}{\theta}

\newcommand{\vi}{\varphi}

\newcommand{\la}{\lambda}

\newcommand{\Sa}{\Sigma}

\newcommand{\ra}{\hat{r}}
\newcommand{\ro}{\hat{r_0}}
\newcommand{\lo}{\hat{L}}
\newcommand{\co}{c}
\newcommand{\di}{d}

\newcommand{\kinu} {\lp u^{\al}\phi_{,\al} \rp}
\newcommand{\p}{p^\phi}
\newcommand{\Phih}{\hat{\Phi}_{\ell m}}

\usepackage{ulem}

\usepackage{color}
\usepackage{subcaption}

\begin{document}

\title{Stability of disformally coupled accretion disks}
\preprint{NORDITA-2015-28} \preprint{ HIP-2015-10/TH}
\author{Tomi S. Koivisto}
\email{tomik@astro.uio.no}
\author{Hannu J. Nyrhinen}
\email{hannu.nyrhinen@helsinki.fi}
\affiliation{Nordita, KTH Royal Institute of Technology and Stockholm University, Roslagstullsbacken 23, SE-10691 Stockholm, Sweden}
\affiliation{Helsinki Institute of Physics, P.O. Box 64, FIN-00014 Helsinki, Finland}
\affiliation{Department of Physical Sciences, Helsinki University, P.O. Box 64, FIN-00014 Helsinki, Finland}

\date{\today}

\begin{abstract}
 
The no-hair theorem postulates that the only externally observable properties of a black hole are its mass, its electric charge, and its angular momentum. In scalar-tensor theories  of gravity, a matter distribution around a black hole can lead to the so called ''spontaneous scalarisation'' instability that  triggers the development of scalar hair. In the Brans-Dicke type theories, this effect can be understood as a result of tachyonic effective mass of the scalar field. 
Here we consider the instability in the generalised  {class of scalar-theories} that feature non-conformal, i.e. ''disformal'', couplings to matter.  {Such theories have gained considerable interest in the recent years and have been studied in a wide variety of systems, both cosmological and astrophysical.} In view of the prospects of gravitational wave astronomy, it is relevant to explore the implications of the theories in the strong-gravity regime.  {In this article, we concentrate on the spontaneous scalarisation of} matter configurations around Schwarzschild and Kerr black holes.  {We find} that in the more generic scalar-tensor theories, the instability of the Brans-Dicke theory can be enhanced, suggesting violations of the no-hair theorem. On the other hand,  {we find that}, especially if the coupling is very strong, or if the gradients in the matter distribution are negligible, the disformal coupling  {tends} to stabilise the system. 

\end{abstract}

\maketitle

\section{Introduction}

Until recently, Einstein's theory of gravitation had been stringently constrained only in the regime where gravity is weak and velocities are small. The detection of gravitational waves  by the LIGO/Virgo collaboration \cite{Abbott:2016nmj} can be regarded as the first experimental verification of the predictions of the theory in the nonlinear, highly dynamical regime. The detected gravitational waves were generated during a collision and merger of two black holes. The environment of such an event is quite complementary to the quasi-static, quasi-linear regime of the classic tests of the gravitational theory, the physics of the Solar System and of binary pulsars \cite{Will:2014kxa}. Due to ongoing investments in gravitational wave astronomy \cite{Somiya:2011np,Aasi:2013wya,AmaroSeoane:2012km}, in the future such events will be observed with increasing precision. This invites to explore the implications of generalised gravitational theories to the astrophysical phenomena characterised by strong gravitational field strengths\footnote{The Newtonian gravitational potential $\sim M/r$, where $M$ is the mass of the source and $r$ the characteristic distance scale. By the virial theorem, the typical velocities are  of the order $v \sim \sqrt{M/r}$. More properly one refers to the curvature, $R^r_{\phantom{r}0r0} \sim M/r^3$, which can be computed as the gauge field strength of the gravitational connection.}.

Quite generically, extensions of General Relativity introduce additional degrees of freedom. In this paper, we restrict to the simplest possible, which is to allow one additional scalar field. In cosmology, models abound where (more and less ad hoc) scalar fields are responsible for the inflation in the early universe, the acceleration of the present universe, or even the apparent need for dark matter. Besides those phenomenological motivations to invoke scalar fields, they are almost ubiquitous in high-energy physics beyond the standard model. In particular, low energy limits of string theory obtained by the plethora of possible compactifications typically involve the dilaton and other moduli that may come with non-minimal gravitational couplings\footnote{As a simplified example, we recently considered a Dirac-Born-Infeld brane filled with vector field matter \cite{Koivisto:2015vda}. The location of the brane is effectively described by a scalar field, and the coupling of the vector field is through the metric of the form (\ref{disformal1}).}. Another example is in the context of supergravitational or other generalised gauge geometries in four dimensions where often the difficulty is to hide away, rather than to predict, new fields (see \cite{Gaillard:2015rpc} for a brief review of phenomenological aspects of superstrings and supergravity). 

Models based on such ideas can now be subjected to new experimental constraints, inasmuch  as  observable deviations of the models from General Relativity are understood in the strong-field regime. Lacking a single compelling alternative theory, it is of interest to uncover the possible phenomenology in generic theoretically consistent models.

It has been well established that the presence of a scalar field $\phi$ can lead to an effect called spontaneous scalarisation of compact objects \cite{Damour:1993hw,Damour:1996ke}. Assuming the field $\phi$ is coupled to the metric curvature, $R$, by the term $\phi R$, the field can develop nontrivial ''hair'' around the massive object sourcing the metric $g_{\mu\nu}$. The appearance of the instability stems from the non-minimal coupling that induces a tachyonic effective mass, that is a negative effective mass squared, for the field\footnote{Another possible phenomenon is that of superradiance \cite{Brito:2015oca}, where the effective mass squared remains positive and exceeds a threshold given by the rotation of the black hole.}.
As a result of the scalarisation, the mass of the object can appear drastically different than in the absence of the scalar field.
Considering the example of a binary merger, the spontaneous scalarisation can have an impact on the dynamics, also beyond dressing the parameters of the system. The consequences are potentially observable especially in the final cycles of the merger, when the field strength is the strongest \cite{Barausse:2012da}. 

Contrary to compact objects such as neutron stars, in the case of black holes, the scalarisation can only occur if there is matter surrounding the black hole. The scalarisation happens if the scalar can settle into a stable configuration instead of radiating away as required by the no-hair theorem. The no-hair theorem postulates that the only externally observable properties of a black hole are its mass, its electric charge, and its angular momentum. In particular, this leads one to consider  the stability or instability of the matter configurations in view of spontaneous scalarisation. The outcome of the scalarisation can be a non-trivial configuration of the scalar field, i.e. scalar black hole hair.  Occurrence of this highly nonlinear effect can be used to constrain deviations from General Relativity in the form of effectively non-minimal couplings. 

 
Models based on the metric tensor $g_{\mu\nu}$ and a scalar field $\phi$ are called scalar-tensor theories of gravity. The new spin-0 interaction they may feature due to the field $\phi$ is known as the fifth force \cite{Will:2014kxa,Joyce:2016vqv}. Those theories are often written in a frame where the scalar field couples non-minimally to gravity, for instance in the form $\phi R$ mentioned above, whereas the matter sector is retained as in General Relativity. The most extensively studied example is the Brans-Dicke theory, where the coupling basically promotes the Newton's constant in the Einstein-Hilbert Lagrangian into a dynamical scalar field. 

The Brans-Dicke theory may also be interpreted as General Relativity sourced by a scalar field that has non-minimal couplings to other matter sources. This is famously seen by a mapping of the theory via the rescaling of the metric, $\hat{g}_{\mu\nu}=\phi g_{\mu\nu}$ \cite{Fujii:2003pa,Capozziello:2011et}. This mapping leads to non-minimally coupled matter sector. In the $\hat{g}_{\mu\nu}$-frame matter does not follow the geodesics of the metric $g_{\mu\nu}$, because the non-minimal coupling to the scalar field, the fifth force, may distort the paths of matter fields. Taking properly into account the rescaling of the physical units, the two frames are classically equivalent \cite{Brans:1961sx}. In particular, such conclusions as whether or not the theory is stable, are independent of in which frame they are interpreted. For clarity, from now on we will  continue the discussion in the ''Einstein frame'' wherein matter (and not geometry) is non-minimally coupled the scalar field $\phi$. 

The non-minimal coupling is essential in the generation of the tachyonic effective mass that triggers the process of scalarisation. A bare potential of the field is not crucial for the mechanism, though we need it to justify the nearly homogeneous scalar background. The kinetic term of the field is of course relevant to the dynamics, and in this paper we permit non-canonical kinetic fields for generality. We find it most interesting however, to extend the framework by incorporating more generic varieties of consistent scalar field couplings. The calculation we undertake in this paper is the (linear) stability of a (homogeneous) scalar field in a strong spherical and axisymmetric gravitational fields. 

Recently, many interesting findings have resulted from explorations into the theory space of more general scalar-tensor gravity. The most general scalar-tensor theories with second order equations of motion are known as the Horndeski theories \cite{horndeski1974second}. In addition to a purely conformal transformation between different frames, given above by $\phi$ and more generally by some function $C$, it is understood that scalar-tensor theories generically feature also a so called disformal piece $D$. Matter follows geodesics of a metric $\hat{g}_{\mu\nu}$, which is related to the gravitational metric ${g}_{\mu\nu}$ as
\be \label{disformal1}
\hat{g}_{\mu\nu}=C(\phi,X)g_{\mu\nu}+D(\phi,X)\phi_{,\mu}\phi_{,\nu}\,.
\ee
Here $\phi$ is the scalar field, $\phi_{,\mu}$ its partial derivatives, and $X=-(\partial\phi)^2/2$ its kinetic term. This is the most general transformation between frames that ''respects the weak equivalence principle, ordinary notions of causality, and which is insensitive to a change of zero for the auxiliary scalar field'' \cite{bekenstein1993}. Indeed, it has been recently clarified that this transformation brings generic classes of Horndeski theories into their Einstein frame \cite{Zumalacarregui:2012us, Bettoni:2013diz} when it exists \cite{Zumalacarregui:2013pma,Bettoni:2015wta}. In cosmology, disformal fields have been applied to models of inflation \cite{Kaloper:2003yf,Koivisto:2014gia}, dark energy \cite{Koivisto:2008ak,Zumalacarregui:2010wj}, interacting dark matter \cite{Koivisto:2012za,vandeBruck:2015ida} and distortions of the cosmic microwave background \cite{vandeBruck:2013yxa,Brax:2013nsa}. In astrophysics, they have been studied in the context of solar system observations \cite{Ip:2015qsa} and neutron stars \cite{Minamitsuji:2016hkk}. Currently, collider physics seems to provide the most stringent constraints on the possible couplings of the standard model particles \cite{Brax:2014vva}. On the other hand, disformal interactions arise naturally exclusively in the dark sector in higher dimensional theories where matter resides upon moving branes \cite{KOIVISTO:2013jwa,Koivisto:2013fta}.


Here we are concerned with astrophysical situations where disformally coupled matter (dark or otherwise) surrounds massive sources, such as black holes. 
The structure of the paper is as follows. In section \ref{set-up} we introduce the covariant system of equations for the set-up we are considering, and perturb them linearly around a static background configuration. In section \ref{shole}, we then analyse the case of a Schwarzschild black hole testing the robustness of our conclusions by considering slightly different profiles for the surrounding matter. In section \ref{khole}, we investigate the effects of rotation by repeating the computations, with some additional approximations, for a Kerr black hole. Conclusions are then summarised in section \ref{conclusions}.

\section{The general set-up}
\label{set-up}

We consider a gravitational system with  perfect fluid matter and a scalar field $\phi$.  {In this system matter couples to a metric} $\hat{g}_{\mu\nu}$  {which} is disformally related to the gravitational metric $g_{\mu\nu}$. We will assume the relation (\ref{disformal1}) only depends upon the scalar field and its derivatives, such that 
\be \label{disformal}
\hat{g}_{\mu\nu}=C(\phi)g_{\mu\nu}+D(\phi)\phi_{,\mu}\phi_{,\nu}\,.
\ee
The {Einstein frame} action we consider is then
\be \label{action}
S=\int \mathrm{d}^4x \sqrt{-g}\lb \frac{R}{16\pi G} + \p(\phi,X)\rb + S_m(\Psi,\hat{g}_{\mu\nu})\,,
\ee
where $\Psi$ denotes the matter fields coupled to the metric (\ref{disformal})  {and $\p(\phi,X)$ consists of potential and kinetic terms of $\phi$}.
The stress energy for the matter content we assume to take the perfect fluid form
\be
T_{\mu\nu} = (\rho + p)u_\mu u_ \nu + pg_{\mu\nu}\,,
\ee
and the stress energy tensor for the field $\phi$ is
\be \label{step}
T^{\phi}_{\mu\nu} = \p g_{\mu\nu} + \p_{,X}\phi_{,\mu}\phi_{,\nu}\,.
\ee
Here we recall that $X$ is the kinetic term $X=-(\partial\phi)^2/2$  and for a canonic scalar field $\p=X-V$. Such cases have been studied in the literature with the perfect fluid taken to be dark matter, baryons, or radiation, see e.g. \cite{Koivisto:2012za,vandeBruck:2015ida,vandeBruck:2013yxa,Brax:2013nsa,Sakstein:2014isa,Sakstein:2014aca}. 

Interestingly, an effective four dimensional action of the form (\ref{action})  {appears for example} in DBI brane scenarios in type II A string theory.  {There} disformally coupled matter is identified with fields residing upon the moving brane \cite{KOIVISTO:2013jwa,Koivisto:2013fta,Koivisto:2014gia} and the functions $C \sim 1/D \sim h^{-\frac{1}{2}}$, where $h(\phi)$ is the warp
factor of the higher dimensional geometry. In that case the Lagrangian for the scalar field has the DBI form $\p = (1-\gamma^{-1})/h - V$, where $\gamma = (1-2hX)^{-\frac{1}{2}}$ is the
Lorentz factor for the movement of the brane. In the following we will use the generalised definition
\be
\ga \equiv \lp 1-2 \frac{D}{C}X \rp^{-\frac{1}{2}}\,.
\ee
which is convenient to exploit also outside the extra dimensional context.

The field equations have the usual form
\be
G_{\mu\nu} = \kappa^2\lp T_{\mu\nu} + T^{\phi}_{\mu\nu}\rp\,,
\ee
The conservation of the stress energies is given by 
\be \label{cons}
\nabla^{\mu} T_{\mu\nu} = -Q\phi_{,\nu}\,, \quad \nabla^{\mu} T_{\mu\nu}^\phi = Q\phi_{,\nu}\,,
\ee
 {where} the coupling function $Q$ can be determined from the general formula \cite{Koivisto:2012za}
\be
Q=-\frac{1}{2C}\lp C'T + D'T^{\mu\nu}\phi_{,\mu}\phi_{,\nu}\rp + \nabla_\mu \lp \frac{D}{C}T^{\mu\nu}\phi_{,\nu}\rp\,.
\ee
Now,  by noting that
\be
\gamma^{-2}Q = \frac{C'}{2C}\lp\rho-3p\rp + \lp\frac{D'}{2C} - \frac{DC'}{C^2}\rp \lb \lp\rho+p\rp \kinu^2 - 2pX\rb + \frac{D}{C}\lb \lp\rho+p\rp u^\mu u^\nu \nabla_{\mu}\phi_{,\nu} + p \Box\phi\rb\, ,
\ee
it becomes evident that the disformal coupling has the nontrivial consequence of mixing the higher order derivatives. 

Finally, we need the Klein-Gordon equation for the field, 
\be \label{kg}
\p_{,X} \Box\phi - \p_{,XX}\phi^{\al}\phi^{\bt}\nabla_\al\phi_{,\bt}-2Xp_{,X\phi} X + p_{,\phi} = Q\,,
\ee
to complete our system of equations.

\subsection{Perturbations of the field}

We shall consider fluctuations of the field around its (covariantly) constant background value. For this purpose, we introduce a perturbation $\varphi$ of the field about the background value $\phi_0$  {such that} 
\be
\phi = \phi_0 + \varphi\,.
\ee
As long as the functions are analytical around the background value of the field we can write
\be
V(\phi) = \sum_{n=-\infty}^\infty V_n \vi^n\,, \quad
C(\phi) = \sum_{n=-\infty}^\infty C_n \vi^n\,, \quad
D(\phi) = \sum_{n=-\infty}^\infty D_n \vi^n\,.
\ee
In the following we will assume that the field has the canonical form $\p=X-V$. This would also  {be} a good approximation for the DBI case when the field is rolling at nonrelativistic speeds {,} such that $\gamma \approx 1$.  Another assumption we shall make, that considerably simplifies the treatment, is that the background field value is a constant, that is $\nabla_\mu \phi_0 = 0$.  Then,  it is easy to see from the Klein-Gordon equation (\ref{kg})  that at the background level 
\be 
V_1 + \frac{C_1}{2C_0}(\rho-3p)=0\,.
\ee
If matter is homogeneously distributed, this is consistent with the field having settled to a nontrivial minimum of its effective potential. Otherwise, we should require $V_1=C_1=0$. This is not an essential restriction to us,  {because from previous studies of fifth forces the conformal coupling $C_1/C_0$ is known to be tightly bound. Moreover,} here we  {focus} on the novel effects due to the disformal coupling $D$. Given negligible contribution of the conformal coupling to the background, it is natural to take the background value of the field to correspond to a minimum of its potential such that $V_1=0$.

With these assumptions, the linear part of the Klein-Gordon equation (\ref{kg}) becomes
\be \label{pkg}
\lp 1-\frac{D_0}{C_0}p\rp\Box\vi - \frac{D_0}{C_0}\lp\rho +p\rp u^\mu u^\nu \nabla_{\mu}\vi_{,\nu} = \lb 2V_2 - \lp\frac{C_1^2}{2C_0^2}-\frac{C_2}{C_0}\rp\lp\rho-3p\rp\rb\vi \equiv V''_{eff}\vi\,.
\ee
From this we already see that the disformal part of the coupling is ineffective unless the system has non-negligible matter flows, pressure, or the field develops gradients. This is a manifestation of the ''disformal screening'' \cite{Koivisto:2012za}.

\section{Schwarzschild black hole}
\label{shole}

First, we shall consider static, spherically symmetric spacetimes. A general ansatz for such  {a metric} is 
\be
ds^2 = -e^{A(r)}dt^2 + e^{B(r)}dr^2 + r^2d\Omega^2\,.
\ee  
Because the metric is static, the fluid can be taken to be at rest.  {That means its properly normalised four velocity is $u^\mu =e^{-\frac{1}{2}A}\delta^\mu_0$. While this assumption restricts our investigation to nonrelativistic matter, it still enables us to study the effects of the disformal coupling to the first order of approximation. }

It is then straightforward to see that the derivative terms in (\ref{pkg}) turn out as 
\ba
\Box\vi & = & -e^{-A}\ddot{\vi} +  e^{-B} \lp \vi'' +\frac{2}{r}\vi' + \frac{1}{2}(A'-B')\vi'\rp + \frac{1}{r^2}\nabla^2_\Omega \vi\,, \\
u^\mu u^\nu \nabla_{\mu}\vi_{,\nu} & = & e^{-A}\ddot{\vi} - \frac{1}{2}e^{-B}A'\vi'\,.
\ea

For the following calculation, it is useful to expand the field in terms of frequency modes and spherical harmonics as
\be
\vi(t,r,\Omega) =  \sum_{\ell, m} \Phi_{\ell m}(r)e^{-i\omega t} Y_{\ell m}(\Omega)\,.
\ee
Equation (\ref{pkg}) then implies the set of equations 
\be \label{modes}
a(r)\Phi_{\ell m}''(r) + b(r)\Phi_{\ell m}'(r) + \lb\omega^2-U(\ell, r)\rb \Phi_{\ell m}(r) = 0\,,
\ee
where the coefficients are given by
\ba
a(r) &=&  e^{A-B}\frac{C_0-D_0p}{C_0+D_0\rho}\,, \\
b(r) &=& a(r)\lb \frac{2}{r}+\frac{1}{2}(A'-B') + \frac{D_0(\rho+p)}{2(C_0-D_0p)}A'\rb\,, \\
U(\ell, r) & = & \frac{a(r)e^{B}}{r^2}\ell(\ell+1) + e^{A}\lp 1+\frac{D_0}{C_0}\rho\rp^{-1}V''_{eff}\,.
\ea
 {To study the properties of the field, we} would like to rewrite equations (\ref{modes}) in the canonical Schr\"odingerian form
\be \label{eq:schr}
\frac{d^2 \hat{\Phi}_{\ell m}}{dR^2} + \lb \omega^2 - \hat{U}(r(R),\ell)\rb \hat{\Phi}_{\ell m} = 0\,,
\ee
 {This}  would allow us to deduce the properties of the solutions from  {a} single potential  $\hat{U}(r(R),\ell)$. For this purpose we need to introduce the
rescaled radial coordinate $R$ and the rescaled mode functions $\hat{\Phi}_{\ell m}$ as follows:
\be \label{newvars}
dr = \sqrt{a(r)}dR\,, \quad \hat{\Phi}_{\ell m} = a^{-\frac{1}{4}}(r)e^{\frac{1}{2}\int \frac{b(r)}{a(r)}dr} \Phi_{\ell m}\,.
\ee 
The details of this transformation are shown in the appendix \ref{sec:coc}. We then obtain  {the effective potential} written in terms of the original radial coordinate $r$, namely
\be
\hat{U}(r,\ell) = U(r, \ell) -\frac{1}{16a}\lp 4aa''-3(a')^2 + 8a'b - 8ab'-4b^2\rp \,.
\ee
A sufficient condition for the occurrence of an instability is that \cite{buell1995potentials}
\be \label{cond}
\int_{2GM}^\infty  {a(r)^{-1/2}} \hat{U}(r,\ell)\mathrm{d}r < 0\,.
\ee
 {It should be noted that the scalarisation may still occur even if the above does not hold. The condition simply states a sufficient condition for the field to develop stable configuration around a black hole. As can be seen e.g. from figure \ref{fig:spotentials} (right) potential wells may develop even if the potential itself remains mainly positive.}

In the following we will study test fluid matter distributions around black holes. Following \cite{Cardoso:2013fwa} we ignore backreaction  {to} matter and simply choose a metric, which in the simplest case is the Schwarzschild one, $e^A=e^{-B}=1-2GM/r$ insert a matter profile $\rho(r)$. In either case  we consider, the shape of $\hat{U}(r,\ell)$ follows then straightforwardly but it takes some algebra. In comparison to \cite{Cardoso:2013fwa}, there is just one extra parameter, $D_0/C_0$, but the potential term is nonetheless more involved. 

\subsection{Constant-density disc of matter}

First, let us study a Schwarzschild black hole with surrounding constant density distribution of matter. Let us thus set $e^A=e^{-B}=1-2GM/r$. Since we are interested in the effects of the coupling, we assume the field to be light enough so that we can ignore the mass $V_2=0$. 
For concreteness, we consider dust-like matter with no pressure with the profile 
\be \label{profile1}
\rho(r) = \rho_0\lb \Theta(r-r_0) - \Theta(r- (r_0+L) )\rb\,.
\ee
This describes a shell\footnote{We could as well consider a disc of matter around the black hole and the results would stay essentially unchanged. Properly one would consider the integral over all space in (\ref{cond}), and in the case of a finite-width disc, the integral over the polar angle contributes only a fraction of $\pi$ to the prefactor that is inessential to our conclusions.} of constant density $\rho_0$. The inner boundary of the shell is at the radial distance $r_0$ from the centre of the black hole and it has the width $L$. 
The variables defined in (\ref{newvars}) now become
\be
R = \sqrt{1+D_0\rho_0}\lb r + 2GM\log{\lp r-2GM\rp}\rb\,, \quad \hat{\Phi}_{\ell m} = \lp 1 + D_0\rho_0\rp^{\frac{1}{4}}\lp 1-\frac{2GM}{r}\rp^{\frac{D_0\rho_0}{4}}r\,.
\ee
For the matter distribution  {(\ref{profile1})} it is possible to compute the effective potential and even its integral (\ref{cond}) analytically.  Thus, from this point on, it is convenient to express the radius, the distance and the width of the matter shell in units of Schwarzschild radii as
\be
\ra \equiv r/(2GM)\,, \quad \ro \equiv r/(2GM)\,, \quad \lo \equiv L/(2GM)\,.
\ee
Furthermore, we'll use the following dimensionless quantities to denote the conformal and the disformal contributions to the coupling, respectively:
\be \label{dimc}
\co \equiv \lp\frac{C_1^2}{C_0^2}-\frac{C_2}{C_0}\rp G^2M^2\rho_0\,, \quad \di \equiv \frac{D_0}{C_0}\rho_0\,.
\ee
In terms of these variables, the effective potential reads
\ba
\hat{U}(\ra,\ell) & = & 
    \frac{\lp \ra-1\rp\lb 1+ \ell(\ell+1)\ra\rb}{4G^2M^2\ra^4}  \quad  \text{outside the shell} \\ \label{inside}
\hat{U}(\ra,\ell)  & = &
    \frac{16\lp \ra-1\rp\lb 1+ \ell(\ell+1)\ra - 2\co \ra^3\rb + \di^2}{64G^2M^2\lp 1+\di \rp\ra^4}  \quad  \text{inside the shell} 
\ea
It is then easy to see effect of the coupling in the following two limits:
\begin{itemize}
\item Purely conformal case.
Let us first consider the purely conformal case $d=0$.  {In this case we} obtain from the integral (\ref{cond}) that an instability occurs when
\be
c > \frac{1+2\ell\lp \ell+1\rp}{4\lo}\,.
\ee
This is in accordance with  previous results for conformally coupled scalar fields \cite{Cardoso:2013opa}.
\item Purely disformal case.
When $c=0$ but $d \neq 0$ the integral (\ref{cond}) becomes more complicated. However, one can directly observe from  {the effective potential} (\ref{inside}) that, as long as $d>-1$, the effect of the disformal coupling is stabilising. 
\end{itemize}

As an explicit example we again study the DBI brane scenarios. Let us consider the extra-dimensional set-up where the coupling arises due to the matter living on the induced metric of a brane.  {As before, the warp factor $ h^{-\frac{1}{2}} \sim C \sim 1/D $.}  In the case that the warp factor has the power-law form $h(\phi) = m^{2n-4}\phi^{-2n}$, where $m$ is the relevant energy scale, the dimensionless coupling functions (\ref{dimc}) become
\be
\co=\frac{1}{2}n(n+1)\frac{(GM)^2\rho_0}{\phi_0^2}\,, \quad \di=\frac{m^{2n-4}\rho_0}{\phi_0^{2n}}\,.
\ee  
The conformal contribution vanishes in the case of constant or quadratic warp, and only between these two cases the effect of $\co$ is stabilising. The case of adS geometry corresponds to $n=1$.  
When $(GMm)^2(\phi_0/m)^{2n-2} \ll 1$ only the disformal part is relevant, and in the opposite limit the conformal one.
An exponential warp factor $h(\phi) = m^{-4}e^{\la\phi/m}$ results in
\be
\co=\frac{1}{8}\la^2\lp\frac{GM}{m}\rp^2\rho_0\,, \quad \di = \frac{e^{\la\phi_0/m}}{m^4}\rho_0\,.
\ee
The effect of $\co$ is always destabilising, and of $\di$ always stabilising.

\subsection{Matter profile with gradients}

\begin{figure}
\begin{center}
\includegraphics[width=0.45\columnwidth]{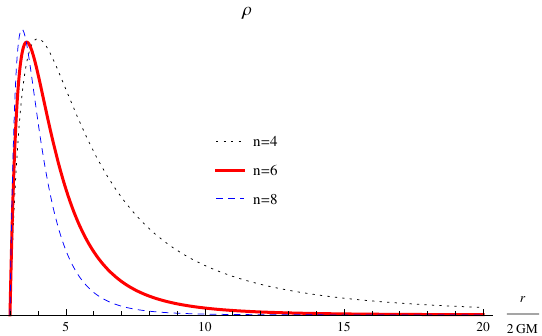} 
\caption{\label{density} The density distribution (\ref{profile1}) when $\ro=3$. }
\end{center}
\end{figure}

 {Next, we move on to study a perhaps more realistic setup, namely a set of density distributions with gradients.} To check whether our conclusions are robust to relaxing  {the simplifying assumption of constant density distribution}, we consider a matter profile of the form
\be \label{profile2}
\rho(r) =\rho_0\Theta(r-r_0)(GM)^{n-1}\frac{r-r_0}{r^n},  {\quad {(n\in\mathbb{N}})}\,.
\ee
As before, this choice  models the existence of an innermost stable circular  {orbit} by not letting matter closer than $r_0$.  {In} addition, the distribution now decays smoothly to zero at large distances given large enough $n$.  {Figure \ref{density} illustrates} the form of the distribution for three choices of $n$. Note that  {now} $\rho_0$ is not  the average density but  {rather} gives the order of magnitude of the maximum density $\rho_{max}=n^{-n}((n-1)/\ro)^{n-1}\rho_0$.

\subsubsection{Purely conformal case}

Setting $\di=0$, the effective potential  {obviously remains intact outside the matter distribution. Inside the matter, that is at}  $\ra > \ro$
\be
\hat{U}(\ra,\ell)= \frac{\ra-1}{2^{n+2}\ra^{n+4}(GM)^2}\lb (2\ra)^n\lp 1+ \ell(\ell+1)\ra\rp - 4\co \ra^3\lp \ra-\ro\rp\rb\,. 
\ee
The result is qualitatively the same as in the previous case: a positive conformal contribution, $\co>0$, can contribute  {to destabilise} the configuration, whereas a negative $\co<0$ adds positively to the effective mass squared. Performing the integral (\ref{cond}), we now find that an instability occurs given
\be \label{icond}
\co > \frac{1}{2}\lb 1 + \ell (\ell+1)(n-1)(n-2)\rb (2\ro)^{n-2}\,.
\ee 
This is in accordance with \cite{Cardoso:2013opa}.

Let us now put in some numbers to see whether the instability is relevant for realistic model parameter values and whether such cases are still within the limits of our approximations. 
The total mass for the profile (\ref{profile1}) is
\be
M_0 = \frac{4\pi^2 (GM)^{n-1}r_0^{n-4}\rho_0}{(n-3)(n-4)}\,.
\ee
We  {assume} that the matter density doesn't backreact to the metric and  {that} the Schwarzschild metric is still a good approximation.  {In other words, we assume that} $M_0$ is sufficiently smaller
than $M$. For  {a Schwarzschild black hole},  the innermost stable circular orbit $r_0=6GM$ \cite{iscot}. The requirement $M_0<M$ then implies an upper limit on the density scale $\rho_0$,
\be
\frac{\rho_0}{M_p^4}  < 2^{n+3}3^{n-4}\pi (n-4)(n-3)\lp\frac{M_p}{M}\rp^2\,. 
\ee 
For a black hole of order Solar mass, $(M_p/M)^2 \sim 10^{-76}$,  which gives  {many} orders of magnitude difference to average cosmic density $\rho/M_p^4 \sim 10^{-122}$. From the condition (\ref{icond}), the spontaneous scalarisation happens when the coupling parameters satisfy
\be
M^2_p\lp\frac{C_1^2}{C_0^2}-\frac{C_2}{C_0}\rp > 2^{3+n}3^{n-2}\pi^2\lp\frac{M_p}{M}\rp^2\lp\frac{M_p^4}{\rho_0}\rp > \frac{9\pi}{(n-4)(n-3)}\,.
\ee
 {Regardless of the density and mass scales, this inequality holds easily for coupling parameters below the Planck scale. Thus, the phenomenon is relevant for canonic quintessence models.} In these models the field is generically of the order of Planck mass and the mass is ultra-light at the present epoch.

\subsubsection{Purely disformal case}

\begin{figure}
\includegraphics[width=0.45\columnwidth]{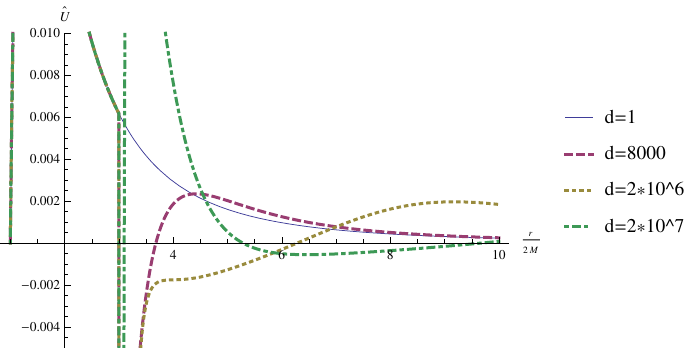}
\includegraphics[width=0.45\columnwidth]{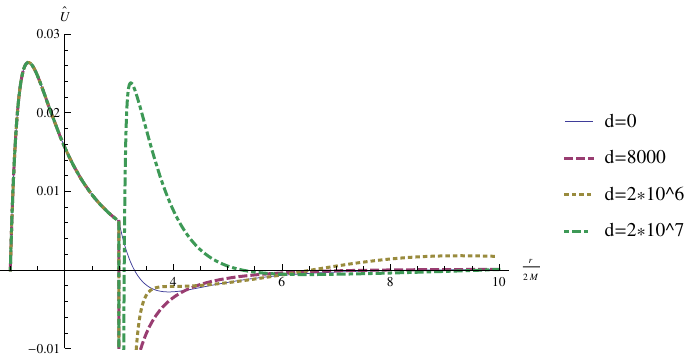}
\caption{Effective potential for a Schwarzschild black hole and for density profile \eqref{profile1} with different disformal coupling strengths. Matter starts at the innermost stable circular orbit ($\ro=3$) and we have chosen $n=6$. The purely disformal contribution (left) can force the potential negative irrespective of the conformal contribution (right). However, large enough disformal coupling adds positively to the potential.} \label{fig:spotentials}
\end{figure}

As the expressions are more complicated in this case, let us consider only the $\ell=0$ angular mode which is the most prone to instabilities.  {Arranged in powers of the disformal coupling factor $\di$,
the effective potential for $\ra>\ro$ reads}
\ba
U(\ra,0) &= & M^{-2} (2 \ra)^{-n-4} \left(2 d (\ra-\ro)+2^n \ra^n\right)^{-3} \Bigg[ 4 (-1 + \ra) (2\ra)^{4 n} + 4 d^4 (\ra - \ro)^4 -  \nonumber \\ 
 && -  d^2 (2\ra)^{2 n} \Big[ \ra^2 \left(32 - 6 n (-3 + \ra) (-1 + \ra) + n^2 (-1 + \ra)^2 + \ra (-38 + 5 \ra)\right) -  \nonumber \\
 &&  - 2 (21 + n (17 + n (-1 + \ra) - 5 \ra) (-1 + \ra) - 22 \ra) \ra \ro + \nonumber \\
 && +  \left(15 - 16 \ra + n (-1 + \ra) (16 - n + (-4 + n) \ra)\right) \ro^2 \Big] - \nonumber \\ 
&& -  4 d^3 (2\ra)^n (\ra - \ro)^2 \Big[2 (-1 + n) \ra^2 + \ro + 2 n \ro + \ra (1 - 2 n (1 + \ro))\Big] + \nonumber \\
&&  +  2 d (-1 + \ra) (2\ra)^{3 n} \Big[(-1 + n) n \ra^2 + (-8 + n (3 + n)) \ro - \nonumber \\
&& -  \ra \left(-10 + n (1 + n) (1 + \ro)\right) \Big] \Bigg].
\ea
The form of the effective potential is illustrated in figure \ref{fig:spotentials}. We see that for a large enough coupling, the potential can have negative values near
the steep rise of density at the distance of the innermost stable circular orbit (which we set as $\ro=3$ corresponding to $r_0=6GM$  {or three Schwarzschild radii} in the figure). Choosing a large enough coupling the integral (\ref{cond}) can indeed go negative, which renders the situation unstable. The spontaneous scalarisation occurs then due to the response of the coupling to the gradient in the matter density profile. The smaller the $n$, the smaller the function $d$ needed for the scalarisation to occur. 

\subsubsection{General coupling}

\begin{figure}[t]
\begin{subfigure}{0.45\textwidth}
\includegraphics[width=\columnwidth]{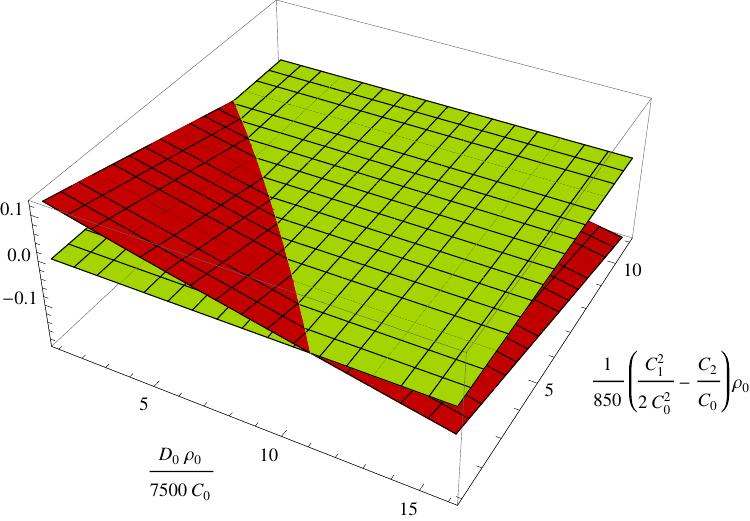}
\caption{The integral $I(c,d)$ for smaller values of $c$ and $d$. For these values the effect of both the conformal and the disformal coupling is destabilising.}
\end{subfigure}
\begin{subfigure}{0.45\textwidth}
\includegraphics[width=\columnwidth]{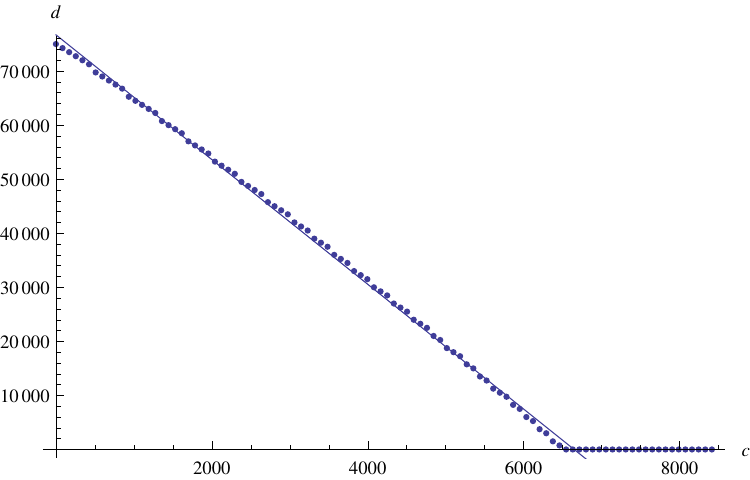}
\caption{Linear fit along $I(c,d) = 0$ in figure (a). Scalarisation occurs above the curve. }
\end{subfigure}
\begin{subfigure}{0.45\textwidth}
\includegraphics[width=\columnwidth]{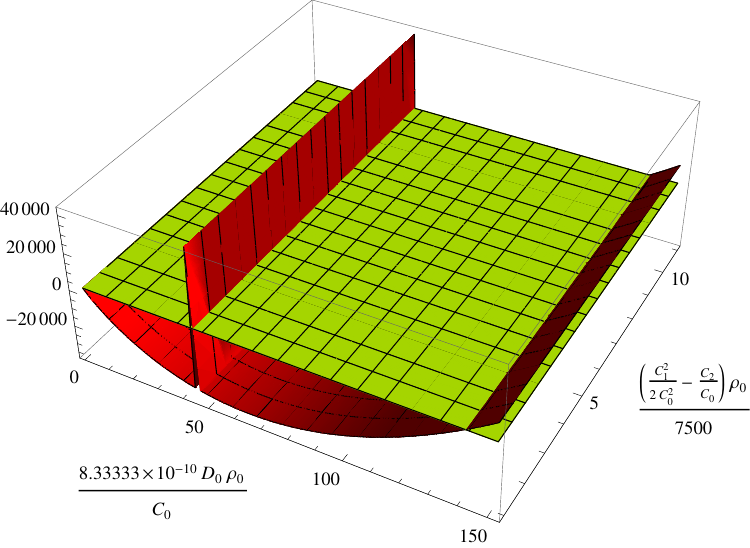}
\caption{Large enough disformal coupling stabilises the integral $I(c,d)$ as expected by the potential in figure \ref{fig:spotentials}.}
\end{subfigure}
\caption{Integral of the effective potential as a function of conformal and disformal couplings $I(c,d)$ ((a) and (c)). We fit linearly along the curve $I(c,d) = 0$ (b) to find a condition for the instability for lower $c$ and $d$.  \label{general}}
\end{figure}

For a general coupling we  {study the condition \eqref{cond} numerically. We integrate the effective potential for the density profile \eqref{profile2} with $n=6$. Figure \ref{general}  shows the integral  at various values of $\co$ and $\di$. We find that also in this case both the conformal and the disformal coupling can contribute to destabilise the system. However, large enough disformal coupling forces the integral positive, thus stabilising the system.

For smaller values of the integral the surface 
\be
I(c,d) \equiv \int_{2GM}^\infty \hat{U}(c,d) dx
\ee 
is nearly flat. Therefore, bearing in mind that we are working at the linear level, the curve $I(c,d) = 0$ can be described by a linear fit. This gives a rough numerical estimate of a condition between the conformal and the disformal coupling for the instability to occur. We find that at linear level the instability condition \eqref{cond} is met when
\be
\di \gtrsim \frac{\ka}{\rho_0} + \la   \co,
\ee
where $\ka = 76500$ and $\la= - 11.5$.  For these values larger coupling further destabilises the system.}

\section{Kerr black hole}
\label{khole}

Next we shall consider a metric of an axially symmetric,  that is a rotating black hole. We use a Boyer-Lindquist type ansatz with one off-diagonal element, namely
\be
ds^2 = -{A(r,\ta)} dt^2 + {G(r,\ta)} dr^2 + \Sa(r,\ta)^2 d\ta^2 + {H(r,\ta)}d\phi^2 - B(r,\ta) dt d\phi 
\ee
where $\Sa^2 = r^2 + a^2 \cos^2 \ta$.

As before we assume the fluid to be static, $u^\mu = A^{-1/2}\delta^\mu_0$, so  the D'Alembertian and the gradient term in \eqref{pkg}  {become} 
\ba
\Box \vi &=& -\frac{H}{A H + B^2} \ddot{\vi} + \frac{1}{G} \vi'' +\frac{G \left(\frac{A' H + A H' + 2 B B'}{A H + B^2}+\frac{2 \Sigma'}{\Sigma }\right)-G'}{2 G^2} \vi'  +\frac{1}{\Sigma ^2} \vi_{,\ta\ta} \nonumber \\
&& +\frac{\frac{\Sigma \left(G \left(A_{,\ta} H+A H_{,\ta} + 2 B B_{,\ta} \right)+G_{,\ta} \left(A  H + B^2 \right) \right)}{G \left(A H + B^2\right)}-2 \Sigma_{,\ta}}{2 \Sigma ^3} \vi_{,\ta} + \frac{A}{A H+B^2} \vi_{,\phi \phi} -\frac{2 B}{A H +B^2}\dot{\vi}_{, \phi} \\
u^\mu u^\nu \nabla_\mu \vi_{,\nu} &=& \frac{1}{A}\ddot{\vi} - \frac{A'}{2 A G} \vi'- \frac{A_{,\ta}}{2 A \Sigma^2}\vi_{,\ta}
\ea
In the case of a rotating black hole the separation of the field to  radial and angular functions is in general not possible. Thus, we investigate the field in the equator ($\theta = \pi/2$) of the black hole where the effect of the rotation is strongest. While this restricts  {the generality of our investigation} it should not be a problem because we are interested in thin disks of matter around black holes. Such disks tend to the equator due to the Bardeen-Petterson effect irrespective of the initial alignment of the disk \cite{bardeen1975}. This also happens in relatively short time  \cite{natarajan1998} and thus we concentrate on the equations at $\ta = \pi / 2$.

On the equator we can use the composition $\vi = \Psi(r) S(\tfrac{\pi}{ 2}) e^{-i \om t + i m \phi}$ where $\om$ is some particular frequency and $m$ the azimuthal number. The Klein-Gordon equation \eqref{pkg} now becomes a pair of equations 
\ba \label{eq:pari}
S''(\tfrac{\pi}{2}) =& 0 \nonumber \\
\al(r)\Phi''(r) + \beta(r)\Phi'(r)+ \lp \omega^2 - U_K(r)\rp \Phi(r) =& 0. 
\ea
Here
\ba
\al(r) &=&  \frac{(2 M-r) \Delta \left(\frac{D_0}{C_0} p\right)}{r \left(\frac{D_0}{C_0} \left(4 a^2 M^2 p+r^2 \rho  \Delta \right)- (2 M-r) \left(a^2 (2 M+r)+r^3\right)\right)} \nonumber \\
\bt(r) &=&\frac{\Delta \left(\frac{D_0}{C_0} \left(p \left(a^2 M-6 M^2 r+7 M r^2-2 r^3\right)+M \rho \Delta \right)+2 r \left(2 M^2-3 M r+r^2\right)\right)}{r^2 \left(\frac{D_0}{C_0} \left(4 a^2 M^2 p+r^2 \rho  \Delta \right)-  (2 M-r) \left(a^2 (2 M+r)+r^3\right)\right)} \\
U_K(r) &=&  \frac{(2 M-r) \left[ \left(r V''_{eff} \Delta+\left(4 a \sqrt{l (l+1)} M \omega + l (l+1) (2 M-r)\right)\right)\left(1-\frac{D_0}{C_0}p \right)\right]}{ (2 M-r) \left(a^2 (2 M+r)+r^3\right)-\frac{D_0}{C_0} \left(4 a^2 M^2 p+r^2 \rho \Delta \right)}\nonumber
\ea
There seems to be no way to get rid of $\omega$ in the potential. Mathematically the frequency dependence is the consequence of the cross-term $dt d\phi$ of the metric (hence the potential also depends on the azimuthal number $m$). It should be noted that each pair of equations \eqref{eq:pari}  thus hold for a specific frequency $\om$.

From here on the investigation proceeds as before and the Shr\"odingerian form \eqref{eq:schr} is obtained using redefinitions 

\be
dr = \sqrt{\al}\, dR \qquad \text{and} \qquad \hat{\Phi} = \alpha^{-\frac{1}{4}}e^{\frac{1}{2} \int \frac{\bt(r)}{\al(r)}dr} \Phi.
\ee
The effective potential we are interested in is now 
\be
\hat{U}_K(r) = U_K(r) - \frac{1}{16 \al }\lp 4 \al \al'' - 3(\al')^2 + 8 \al' \bt -8 \al \bt' - 4 \bt^2 \rp 
\ee

\begin{figure}[]
\includegraphics[width=0.45\columnwidth]{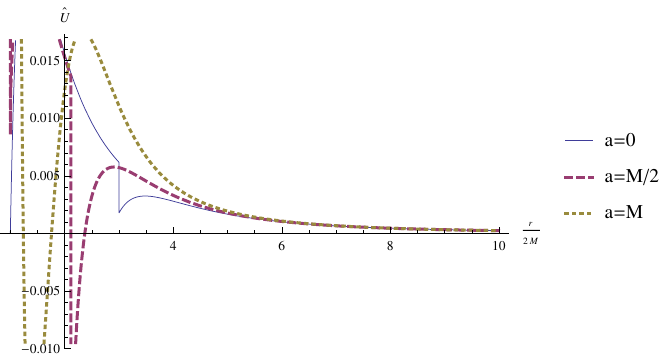}
\includegraphics[width=0.45\columnwidth]{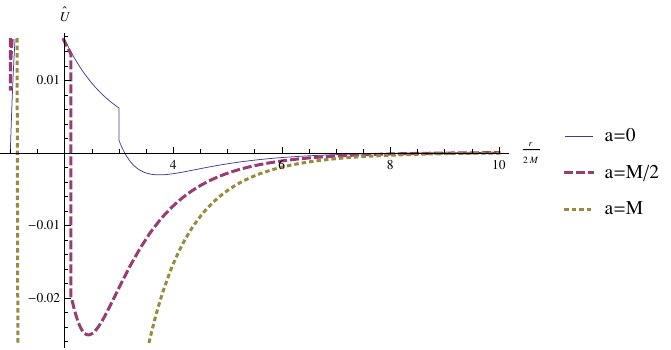}
\caption{For a rotating black hole the potential well grows deeper with angular momentum. That is to say the faster the rotation, the more unstable the configuration. This is true for both the purely disformal coupling (left) as for the general coupling (right). \label{diffKerr}}
\end{figure}

 {The effect of the rotation can be seen in figure \ref{diffKerr}. Firstly, for a rotating black hole the wells are deeper and peaks are higher than for a Schwarzschild black hole. Secondly the wells as well as the peaks shift closer to the black hole.} This shift is due to the fact that we assume the disk to start from the innermost stable circular orbit. For a Kerr black hole the distance of this orbit from the horizon depends on the angular momentum of the black hole \cite{iscot}. 

\begin{figure}[]
\includegraphics[width=0.45\columnwidth]{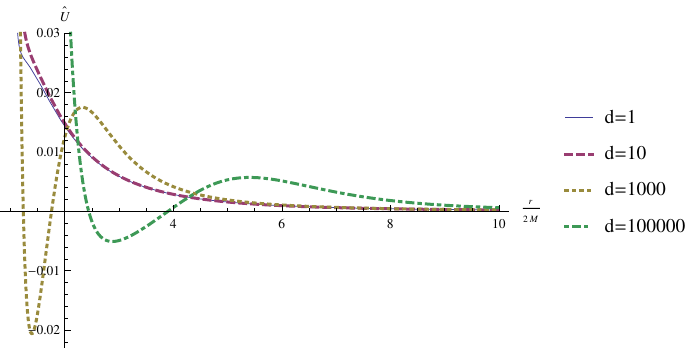}
\includegraphics[width=0.45\columnwidth]{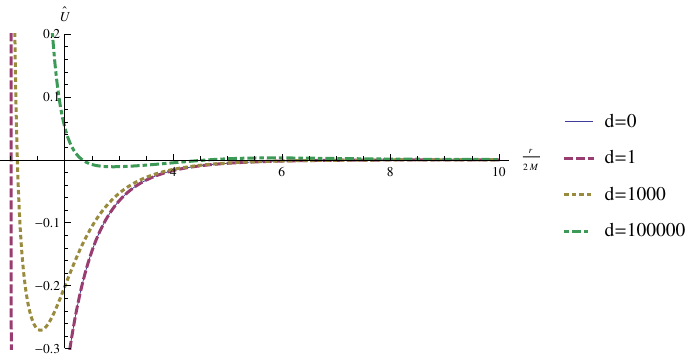}
\caption{For a maximally rotating black hole the disformal coupling seems to stabilise the system with general coupling (right)  {seems to stabilise the system, though our assumptions break down near the ergoregion.} \label{MaxKerr} }
\end{figure}

In figure \ref{MaxKerr} the potential with general coupling (taking into account both the disformal and conformal couplings) is stabilised by the disformal contribution. However, the problems may only be swept under a rug, in this case inside the ergosphere. Near the Schwarzschild radius the velocity of matter on a Keplerian orbit is already 0.7, which makes it questionable whether our assumption of stationary matter is still reasonable. Therefore, we restrict our investigation outside the ergoregion altogether.

\begin{figure}
\includegraphics[width=0.45\columnwidth]{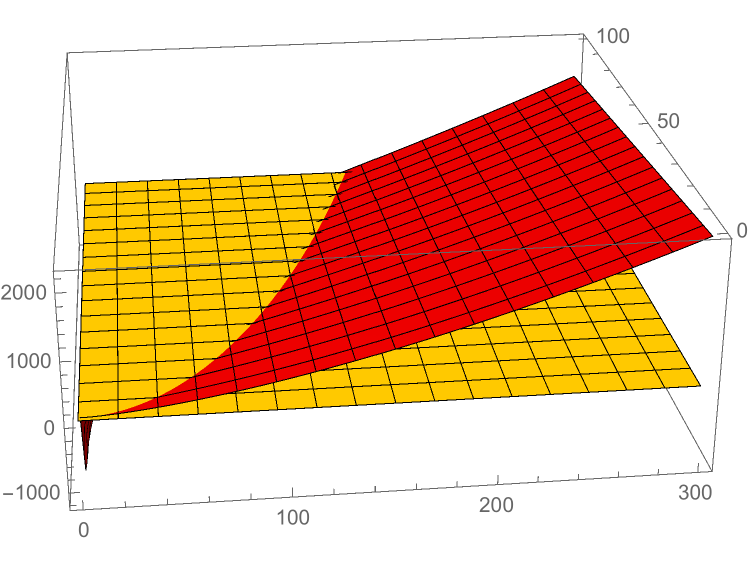}
\caption{For a maximally rotating black hole the disformal coupling seems to stabilise the system with general coupling. The values of the integral of the effective potential increase with the disformal coupling strength. \label{fig:kerr} }
\end{figure}

There is also another difficulty related to exact computation of the integral. That is the fact that the effective potential tends to infinity at the distance of a Schwarzschild radius from the centre and the integral does not converge. However, as can be seen from figure \ref{fig:kerr} in this case the larger the disformal coupling the more positive the integral. Furthermore, it should be noted that the effective potential tend to positive infinity near the Schwarzschild radius. Therefore, the contribution near the ergoregion only adds positively to the integral.

\section{Conclusions}
\label{conclusions}

We studied the phenomenology of generalised scalar-tensor theories, featuring the disformal relation (\ref{disformal}) between the gravitational and matter geometries. In particular, we considered black holes surrounded by matter, such as accretion discs around astrophysical black holes. In such situations, the coupling can induce an instability of the scalar field, that  {triggers} the formation of scalar hair via the spontaneous scalarisation effect. This violates the no-hair theorem, according to which a black hole should not have such externally observable properties in addition to mass, angular momentum, and electric charge.

In our approximation, we considered linear perturbations in the scalar field. We neglected higher order perturbations, and furthermore assumed that the backreaction of the scalar field  {to} the matter distribution and  {to} the background geometry is negligible. These assumptions can be justified when the fluctuations are small and the amount of total mass in the surrounding matter is sufficiently smaller than the mass of the black hole. We verified that these assumptions can be valid for cosmologically relevant scalar-tensor models. Of course however,  {from our analysis  we can only learn about the stability of linear fluctuations}, whereas addressing the problem of final stable configuration of the matter-black hole-scalar field system would require comprehensive nonlinear calculations.

Within these approximations, we were able to analytically understand the effects of disformal coupling to the stability of the scalar fluctuations. We first focused on the simple set-up where the matter configuration around the black hole is in the form of constant-density disc. Then we established that the disformal coupling always adds a positive contribution to the effective mass squared of the scalar field:  that is, it works against the instability, that could be induced by the conformal piece of the coupling. 

We studied also the case when the matter profile has nontrivial gradients, and in this case we found numerically that the disformal part can add to the instability. However, it is not entirely clear how realistic such configurations are. The possible problem is that we assumed the background scalar field to remain constant which might not be well justified in  inhomogeneous media to which the field is explicitly coupled. Nevertheless, this shows that in principle it is possible that the disformal coupling  causes instabilities, if only in the case of gradients in the matter distribution. Furthermore, we considered also Kerr-Newman black holes in order to see the effect of rotation on the instability. We found that rotation of the black hole doesn't change our conclusions qualitatively. 

In the coming era of gravitational wave astronomy, we need better theoretical understanding of modified gravity theories in the strong gravity regime. These theories predict a variety of effects that deviate from General Relativity. Until recently, our only constraints to such deviations come from observing gravitational phenomena in weak fields, such as in the Solar System, or binary pulsars. Increasing number and precision of gravitational wave telescopes finally allows us to test the predicted effects in the environments of strong gravity. Much theoretical work is needed in order to be able to fully take advantage of the information which the future observations will provide us with.

\acknowledgments
We want to thank Dr. Miguel Zumalacarregui for valuable comments on the first version of this manuscript.
HJN is supported by Magnus Ehrnrooth Foundation and Vilho, Yrj\"o and Kalle V\"ais\"al\"a Foundation.

\appendix
\section{Change of coordinates}
\label{sec:coc}

To transform equation \eqref{modes} to \eqref{eq:schr} we need a new field $\Phih$ and a new variable $R$ (not Ricci scalar). To solve the connection between  {the new} and the old variables in equation \eqref{modes} we write
\be
\Phih = y(r) \Phi_{\ell m} \quad \text{and} \quad \frac{d R}{d r} = x(r)
\ee
Next we solve the unknown functions $y(r)$ and $x(r)$ by substitution of the new variables into \eqref{modes}. First of all we have
\begin{subequations}
\ba
\Phi_{\ell m}'(r) &=& -\frac{y'(r)}{y^2} \Phih + \frac{1}{y} \frac{d R}{d r} \Phih'(R), \\[10pt]
\Phi_{\ell m}''(r) &=&  -\lp \frac{y''(r)}{y^2} - \frac{(y')^2}{y^3}\rp \Phih - \lp 2 \frac{y'(r)}{y^2} \frac{d R}{d r} - \frac{1}{y} \frac{d^2 R}{d r^2} \rp \Phih'(R) + \frac{1}{y} \lp \frac{d R}{d r} \rp^2 \Phih''(R),
\ea
\end{subequations}
where $'$ denotes differentiation with respect to the argument. Inserting the above into \eqref{modes} gives 
\ba \label{newfieldeq}
&&\frac{a}{y} \lp \frac{d R}{dr} \rp^2 \Phih''(R) - \frac{a}{y} \lp 2 \frac{y'(r)}{y} \frac{d R}{d r} - \frac{d^2 R}{d r^2} \rp \Phih'(R) - \frac{a}{y} \lp \frac{y''(r)}{y} - \frac{(y')^2}{y^2}\rp \Phih - \nonumber \\
&& - \frac{b}{y} \lp \frac{y'(r)}{y} \Phih - \frac{d R}{d r} \Phih'(R) \rp + \frac{1}{y} \lb \omega^2 - U(\ell,m) \rb \Phih \nonumber \\
&=& \frac{a}{y} \lp \frac{d R}{dr} \rp^2 \Phih''(R) - \lb \frac{a}{y} \lp 2 \frac{y'(r)}{y} \frac{d R}{d r} - \frac{d^2 R}{d r^2}  \rp - \frac{b}{y} \frac{dR}{dr} \rb \Phih'(R) \nonumber \\
&& + \frac{1}{y} \lb\om^2 -U(\ell,m) - a \lp \frac{y''(r)}{y} - \frac{(y')^2}{y^2}\rp - b \frac{y'}{y} \rb \Phih = 0
\ea 
so in order for this to be the Schr\"odingerian equation \eqref{eq:schr} for $\Phih$ we must have
\be
\frac{dR}{dr} = \frac{1}{\sqrt{a(r)}} \quad \text{which also gives} \quad \frac{d^2 R}{dr^2} =-\frac{a'(r)}{2 \sqrt{a^3}}
\ee
and
\ba
a\lp  \frac{2y'}{y \sqrt{a}} + \frac{a'(r)}{2 \sqrt{a^3}} \rp - \frac{b}{\sqrt{a}} = 0 \quad \text{or}\nonumber \\
y' + \frac{a' - 2b}{4a} y =0.
\ea
The answer to the last equation is of course
\be
y(r) = e^{-\int \frac{a' - 2b}{4a} dr } = a^{-\frac{1}{4}} e^{\frac{1}{2}\int \frac{b(r)}{a(r)} dr},
\ee

Substituting the above into \eqref{newfieldeq} gives the potential term as
\be 
\hat{U}(r,\ell) = U (r,\ell) - \frac{1}{16a} \lp 4 aa'' - 3 (a')^2 + 8a'b - 8ab' -4b^2 \rp 
\ee

\bibliography{refs}

\end{document}